\documentclass[a4paper]{article}
\usepackage{amsmath}
\usepackage{epsfig}
\usepackage{enumerate}
\usepackage{amsbsy}
\usepackage{amsfonts}
\usepackage{amssymb}
\usepackage{amstext}
\usepackage{float}
\usepackage[a4paper,margin=2.4cm,footskip=.5cm]{geometry}

\begin{document}

\pretolerance=10000

\title{\bf Measurement of the cosmic ray \\ all-particle and light-component energy spectra \\ with the ARGO-YBJ experiment}
\date{}
\author{ A. D'Amone, I De Mitri and A.Surdo (on behalf of the ARGO-YBJ coll.) }
\maketitle

Dipartimento di Matematica e Fisica ``Ennio De Giorgi'', Universit\`a del Salento, Lecce, Italy 

Istituto Nazionale di Fisica Nucleare, INFN, Sezione di Lecce, Italy

\vskip 1.cm

\abstract{
ARGO-YBJ preliminary results of the measurements of the all-particle and light-component (i.e. protons and helium) 
energy spectra between approximately 5 TeV and 5 PeV are reported and discussed.  
The resulting all-particle spectrum (measured in the energy range 100$\,$TeV - 3$\,$PeV) is in good agreement 
with both theoretical parametrizations and previous measurements.
The light-component (i.e. p + He) has been measured with high resolution up to about 5 PeV. 
The preliminary result is in agreement with direct measurements and then show a clear indication 
of a bending below 1$\,$PeV. 
Improvements of event selection/reconstruction with the full statistics and a complete 
analysis of systematic uncertainties is currently under way.
}

\section{Introduction}

There is a general consensus that galactic cosmic rays (hereafter CRs) up to the ``knee'' of the all-particle spectrum
($\sim 4 \,$PeV) mainly originate in Supernova Remnants (SNRs). 
%The theoretical modelling of this mechanism can reproduce the measured spectra and composition of CRs. 
%
Recent measurements carried out by the balloon-borne CREAM experiment \cite{cream1,cream2} 
show that the proton and helium spectra from 2.5 to 250 TeV are harder compared to
lower energy measurements. 
As pointed out by several authors, the evolution of the proton and helium spectra and their subtle 
differences could be indications of the presence of different populations of CR sources
contributing to the overall flux and operating in environments with different chemical compositions \cite{gst,blasi11}.
Diffusion effects during CR propagation in the Galaxy might also play an important role.

In the knee region (and above) the measurements of the CR primary spectrum are carried out by EAS arrays. 
In this case mass composition studies are extremely difficult and often affected by large systematic uncertainties.
In the standard picture the average composition at the knee is dominated by light elements, and the knee 
itself is interpreted as the steepening of the p and He spectra \cite{kascade}. 
However, several experimental results suggest an heavier composition at knee energies
\cite{aglietta04,macro,tibet,basje,delhad,casamia}
%
%For instance a hybrid measurement has been carried out by the EAS/TOP and MACRO experiments (by detecting, in coincidence,
%EAS Cherenkov light at 2000 m a.s.l. and underground muons below about 3000 m of water equivalent depth, respectively).
%The result implies a decreasing proton contribution to the primary flux well below 
%the observed knee in the primary spectrum \cite{aglietta04}.
%%
%The same indication was previoulsy given by the analysis of the underground muon component alone by the 
%MACRO experiment \cite{macro}.
%In addition, also the results of the Tibet AS$\gamma$ and the BASJE experiments, located at 4300 m a.s.l 
%and at 5200 m a.s.l. respectively, do favour a heavier composition because the proton component 
%is no more dominant at the knee \cite{tibet,basje}.
%Indications for a substantial fraction of nuclei heavier than helium at 1$\,$PeV have also been 
%obtained in measurements with delayed hadrons \cite{delhad}
%as well as by the CASA-MIA collaboration \cite{casamia}.
%

A measurement of the CR primary energy spectrum (all-particle and light-component) in the energy 
range few TeV - 10 PeV is under way with the ARGO-YBJ experiment (for a description of the 
detector and a report of the latest physics results see \cite{ricap13}). 
In order to cover this wide energy range, different approaches have been followed:

\noindent
- \emph{'digital readout'}, based on the RPC readout strip multiplicity, in the 5$\,$TeV - 200$\,$TeV 
range \cite{bartoli12};

\noindent
- \emph{'analog readout'}, based on the particle density near the shower core, 
in the 100$\,$TeV-10$\,$PeV range;

\noindent 
- \emph{'hybrid measurement'}, carried out by ARGO-YBJ and a wide field of view 
Cherenkov telescope, in the 100 TeV - PeV region \cite{argo-wfcta}.

Preliminary results concerning the all-particle and the light-component (i.e. p+He) spectra obtained 
with the analog readout are summarized in the following. More details on the results obtained with the 
'hybrid measurement' are also given in \cite{argo-wfcta,caozhen}.

\section{Measurement of the all-particle spectrum}

The measurement of the CR energy spectrum up to 10 PeV is under way exploiting the RPC charge readout of the ARGO-YBJ 
detector which allows studying the structure of the particle density distribution in the shower core region up to 
particle densities of about 10$^{4}$/m$^2$ \cite{argo-bigpad1}.
%
%
%%%%%%%%%%%%%%%%%%%%%%%%%%%%%%%%%%%%%%%%%%%%%%%%%%%%%%%%%%%
\begin{figure*}
\begin{minipage}[t]{.47\linewidth}
  \centerline{\includegraphics[width=\textwidth, height=6.5cm]{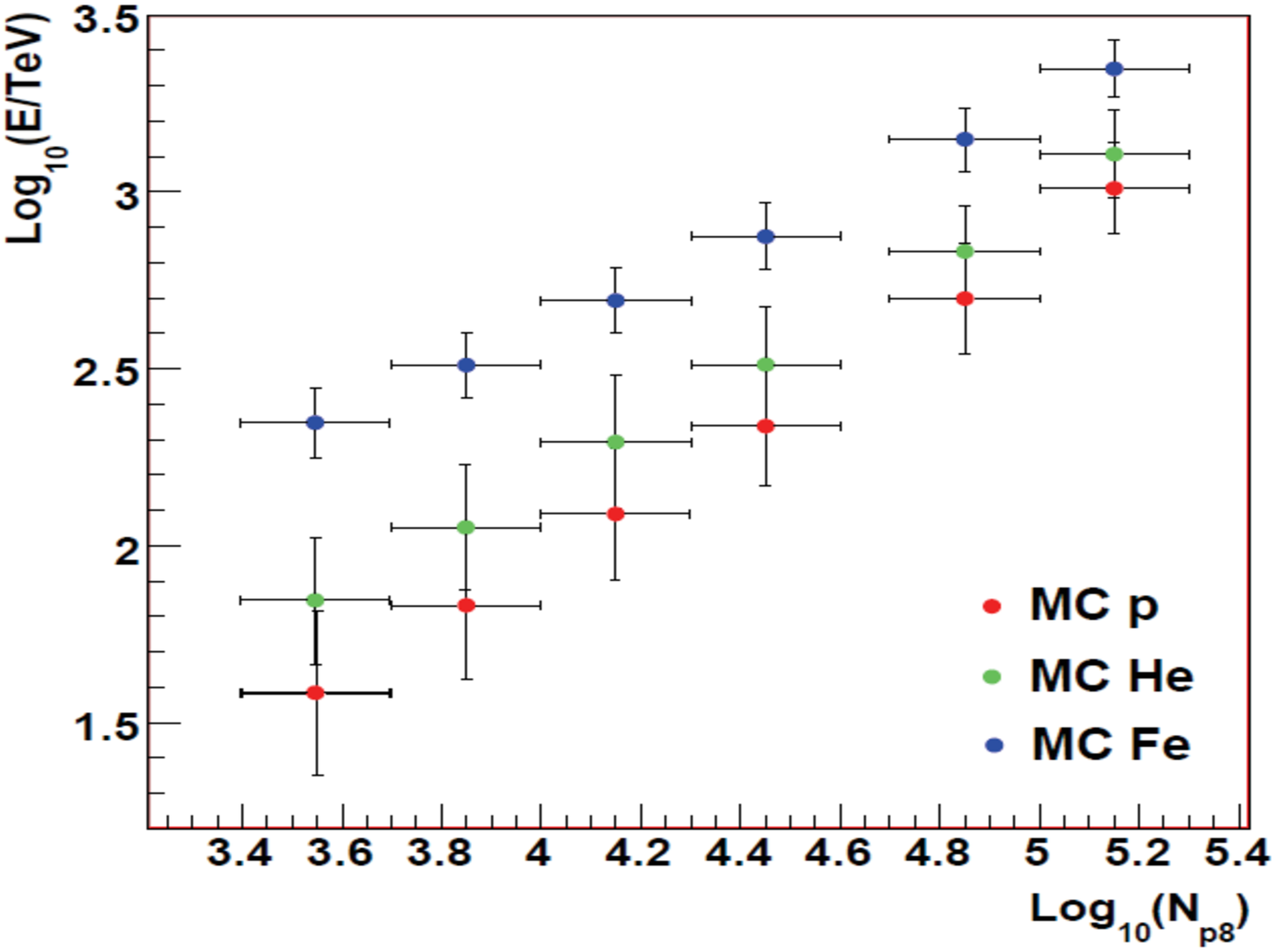} }
\caption[h]{Primary energy as a function of the observed truncated size N$_{p8}$ for simulated showers 
due to different primary nuclei.}
\label{fig:np8}
\end{minipage}\hfill
\begin{minipage}[t]{.47\linewidth}
  \centerline{\includegraphics[width=\textwidth, height=6.8cm]{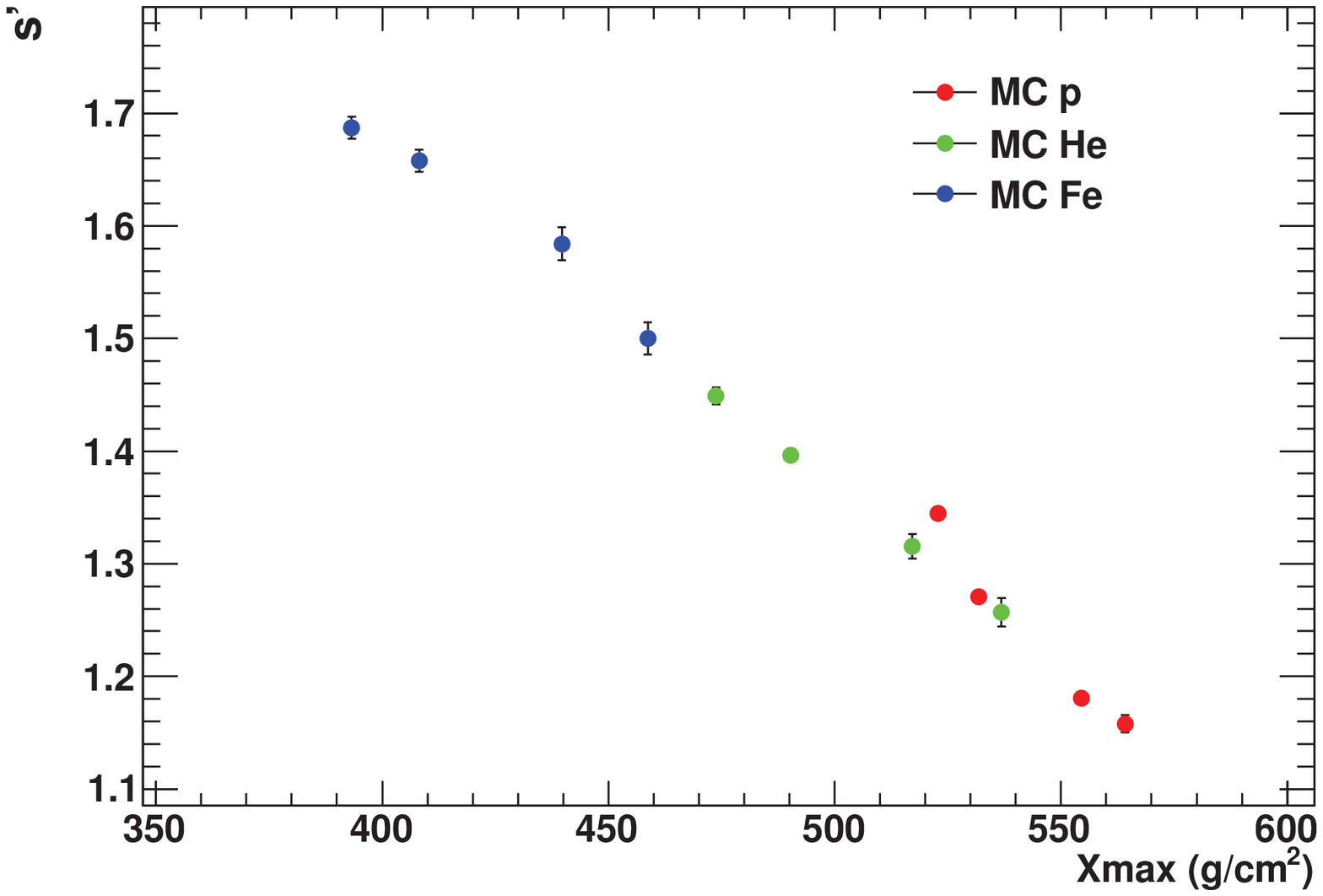} }
\caption{The age parameter $s'$ resulting from the fit of the average LDF for simulated showers vs the corresponding $X_{max}$ 
average values.} 
\label{fig:s_Xmax_pHeFe}
\end{minipage}\hfill
\end{figure*}
%%%%%%%%%%%%%%%%%%%%%%%%%%%%%%%%%%%%%%%%%%%%%%%%%%%%%%%%%%%
%
%
%
%
%%%%%%%%%%%%%%%%%%%%%%%%%%%%%%%%%%%%%%%%%%%%%%%%%%%%%%%%%%
\begin{figure*}
     \vskip -.5cm
     \centerline{\includegraphics[width=0.95\textwidth,height=8.0cm]{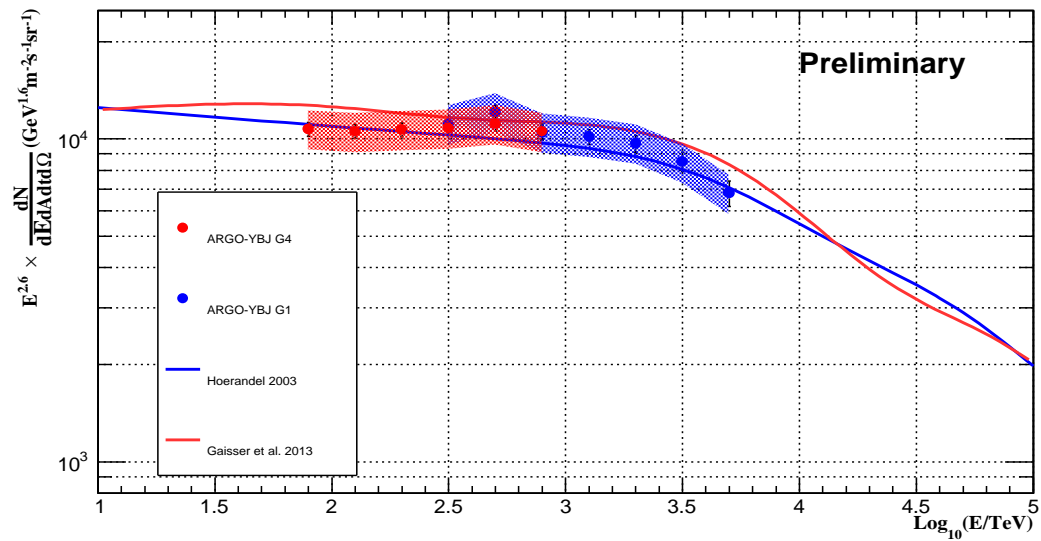} }
\caption{All-particle energy spectrum of primary CRs measured by ARGO-YBJ (see text). 
The parametrizations provided by \cite{horandel} and \cite{gst} are shown for comparison. }
\label{fig:allpart}       % Give a unique label		
\end{figure*}
%%%%%%%%%%%%%%%%%%%%%%%%%%%%%%%%%%%%%%%%%%%%%%%%%%%%%%%%%%%
%
%
The study of the charged particle lateral density function (LDF) at ground is expected to provide information on the longitudinal 
profile of the showers in the atmosphere, that is to estimate their development stage, or the so-called \emph{age}, which is 
related to $X_{max}$, the atmospheric depth at which the cascade reaches its maximum size. 
This implies the possibility of selecting showers within given intervals of $X_{max}$ or, equivalently, 
of $X_{dm}$, the grammage between the depth of the shower maximum and the detection level.
%
% The shower development stage in the atmosphere, as observed at a fixed altitude (the detection one), depends on the
% energy of the interacting primary. For fixed energy, it depends on the nature of the primary: heavy primaries interact
% higher in the atmosphere, thus giving showers which, on average, reach their maximum at a larger distance from 
% the detector than a lighter primary of the same energy. 
%
For this reason, the combined use of the shower energy and age estimations can ensure a sensitivity to the 
primary mass, thus giving the possibility of selecting a light (p+He) event sample with high efficiency.

Various observables were considered and analyzed in order to find a suitable estimator of the primary CR energy. 
Among them, according to MC simulations, N$_{p8}$, the number of particles detected 
within a distance of 8\,m from the shower axis, 
resulted well correlated with energy, not biased by the finite detector size and not much affected 
by shower to shower fluctuations \cite{icrc779}. 
%
% Therefore, the analysis is carried out in terms of 
% different N$_{p8}$ intervals to select event samples corresponding to different primary energies. 
Nevertheless, as shown in Fig. \ref{fig:np8}, this truncated size is a mass-dependent energy estimator parameter.
In order to have a mass-independent parameter we fitted the LDFs of individual showers 
(up to 10\,m from the core) event-by-event, for different N$_{p8}$ intervals and different shower 
initiating primaries, with a suitable function to get the shape parameter $s'$ 
(see \cite{icrc779,icrc781} for details).
From these studies we find that, for a given primary, the $s'$ value decreases when N$_{p8}$ 
(i.e. the energy) increases, this being due to the observation of younger (deeper) showers at larger energies.
Moreover, for a given range of N$_{p8}$, $s'$ increases going from proton to iron, as a consequence of older
(shallower) showers.
Both dependencies are in agreement with the expectations, the slope $s'$ being correlated with the shower age, 
thus reflecting its development stage.
This outcome has two important implications, since the measurements of $s'$ and N$_{p8}$ can both (i) help constraining 
the shower age and (ii) give information on the primary particle nature.

Concerning the first point, we show in Fig. \ref{fig:s_Xmax_pHeFe} the $s'$ values as obtained from the fit of the average LDFs, 
for each simulated primary type and N$_{p8}$ interval, as a function of the corresponding $X_{max}$ average value.
As can be seen, the shape parameter $s'$ depends only on the development stage of the shower, independently from 
the nature of the primary particle.
That plot expresses an important universality of the LDF of detected EAS particles in terms of the lateral shower age. 
The LDF slope $s'$ is a  mass-independent estimator of the average $X_{max}$.
% This also implies the possibility to select most deeply penetrating showers (and quasi-constant $X_{dm}$ intervals) 
% at different zenith angles, an important point for correlating the exponential angular rate distribution with 
% the interaction length of the initiating particle \cite{aielli09}. 
Obviously shower-to-shower fluctuations introduce unavoidable systematics, 
whose effects can be anyway quantified and taken into account.
Another implication is that $s'$ from the LDF fit close to the shower axis, together with the measurement of 
the truncated size N$_{p8}$, can give information on the primary particle nature, 
thus making possible the study of mass composition and the selection of a 
light-component data sample.

Assuming an exponential absorption after the shower maximum, we get N$_{p8}^{max}$, a variable linearly correlated to 
the size at the shower maximum, by using N$_{p8}$ and $s'$ measurements for each event and simply correcting 
with an exponential attenuation: 
$N_{p8}^{max}\approx\ N_{p8}\cdot exp[(h_0 sec\theta - X_{max}(s'))/\lambda_{abs}]$. 
A suitable choice of the absorption lenght $\lambda_{abs}$ (=120 g/cm$^2$) allows to get N$_{p8}^{max}$, 
a parameter correlated with primary energy in an almost linear and mass independent way, 
providing an energy estimator with a Log(E/TeV) resolution of 0.10--0.15 (getting better with energy).

As described in \cite{argo-bigpad1}, the RPC charge readout system has 8 different and overlapping gain scale settings 
(G0,....,G7 from smaller to larger gains) in order to explore the particle density range $\approx$20 -- 10$^4$ particles/m$^2$.
In this preliminary analysis the results obtained with two gain scales (so-called G1 and G4) are presented.
Selecting quasi-vertical events ($\theta$ $<$ 15$^{\circ}$) in terms of the truncated size N$_{p8}$ with the described procedure 
we reconstructed the CR all-particle energy spectrum shown in the Fig.\ref{fig:allpart} in the energy range 100 -- 3000 TeV. 
In the plot a $\pm$14\% systematic uncertainty, due to hadronic interaction models, selection criteria, unfolding algorithms, 
and aperture calculation, is shown by the shaded area. The statistical uncertainty is shown by the error bars.
As can be seen from the figure, the two gain scales overlap making us confident about the event selection and the analysis procedure. 
The ARGO-YBJ all-particle spectrum is in fair agreement with the parametrizations provided by \cite{horandel} 
and \cite{gst}, showing evidence of a spectral index change at an energy consistent with 
the position of the knee. 
As shown in Fig.\ref{fig:all} this result is also consistent with previous measurements made by both direct and indirect experiments.
This is also an important check on the absolute energy scale set for this analysis, whose systematic uncertainty has been anyhow 
conservatively estimated at the level of $10\%$.

\section{Measurement of the light-component energy spectrum}

The CR light-component energy spectrum has been measured by ARGO-YBJ from about 5$\,$TeV to 700$\,$TeV 
\cite{bartoli12,argo-wfcta,ricap13}.
The energy range is now being extended up to the few PeV region by using the RPC charge readout information 
and three different approaches.

\begin{figure*}
\centerline{\includegraphics[width=0.9\textwidth]{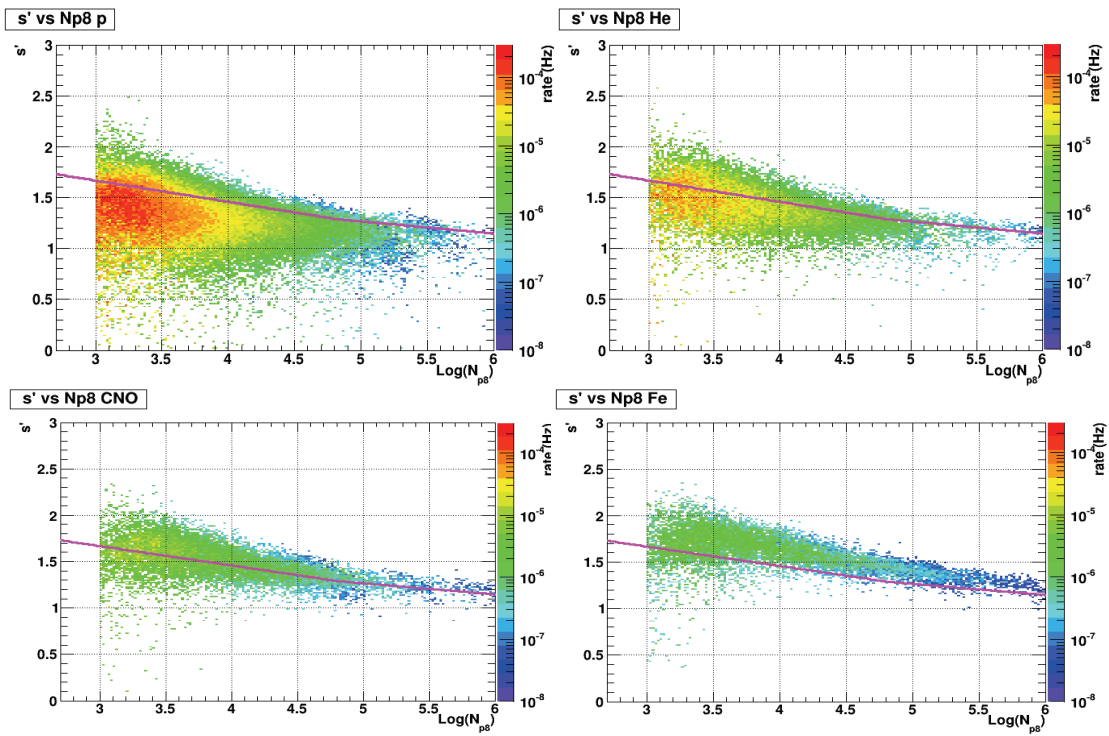} }
\caption{Relation between the LDF shape parameter $s'$ and the truncated size $N_{p8}$ for different nuclei. 
The p+He selection cut is shown by the lines.}
\label{fig:lightsel}       % Give a unique label
\end{figure*}

\begin{itemize}

\item[(1)] A selection of events in the $s'$ -- $N_{p8}$ space allowing to get a light-component sample of showers 
with a contamination of heavier nuclei less than about 15\% (see Fig. \ref{fig:lightsel}). 

\item[(2)] A Bayesian unfolding technique similar to that applied to measure the light-component spectrum 
up to 200 TeV \cite{bartoli12,ricap13,ricap-mm-14}.

\item[(3)] The ARGO-YBJ/WFCTA hybrid measurement \cite{argo-wfcta} 
with a different selection procedure which 
increases the aperture of a factor 2.4, thus allowing the extension to larger energies 
(see \cite{caozhen} for a detailed description of the method and a discussion of the results).
\end{itemize}

\begin{figure*}
\centerline{\includegraphics[width=1.0\textwidth]{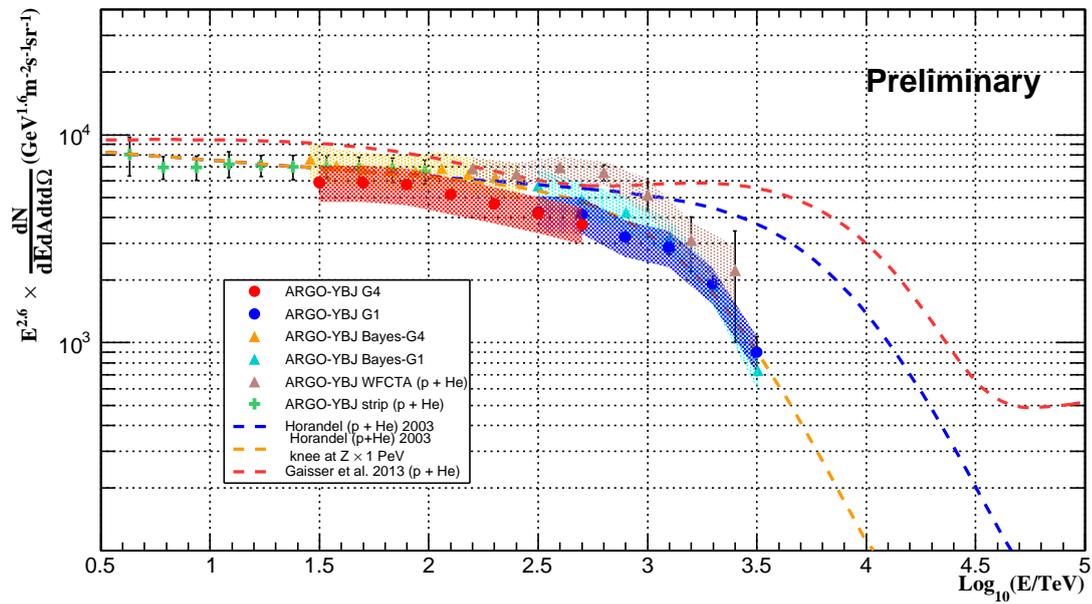} }
\caption{Light (i.e. p+He) component energy spectrum of primary CRs measured by ARGO-YBJ with 
four different analyses (see text).}
\label{fig:phe-knee}       % Give a unique label
\end{figure*}

Preliminary results of the three analyses (and the previous ARGO-YBJ measurement below 200$\,$TeV)
are summarized in Fig.\ref{fig:phe-knee}. 
The systematic uncertainty on the flux is shown by the shaded area and the statistical one by the error bars.
A systematic uncertainty on the energy scale at the level of 5-10\% (depending on the analysis) has also
been conservatively estimated (not shown in the plots).

%%%%%%%%%%%%%%%%%%%%%%%%%%%%%%%%%%%%%%%%%%%%%%%%%%%%%%%%%%
%
%%%%%%%%%%%%%%%%%%%%%%%%%%%%%%%%%%%%%%%%%%%%%%%%%%%%%%%%%%%
%

%
%%%%%%%%%%%%%%%%%%%%%%%%%%%%%%%%%%%%%%%%%%%%%%%%%%%%%%%%%%
\begin{figure*}
\vskip -1.cm
\centerline{\includegraphics[width=1.0\textwidth,clip]{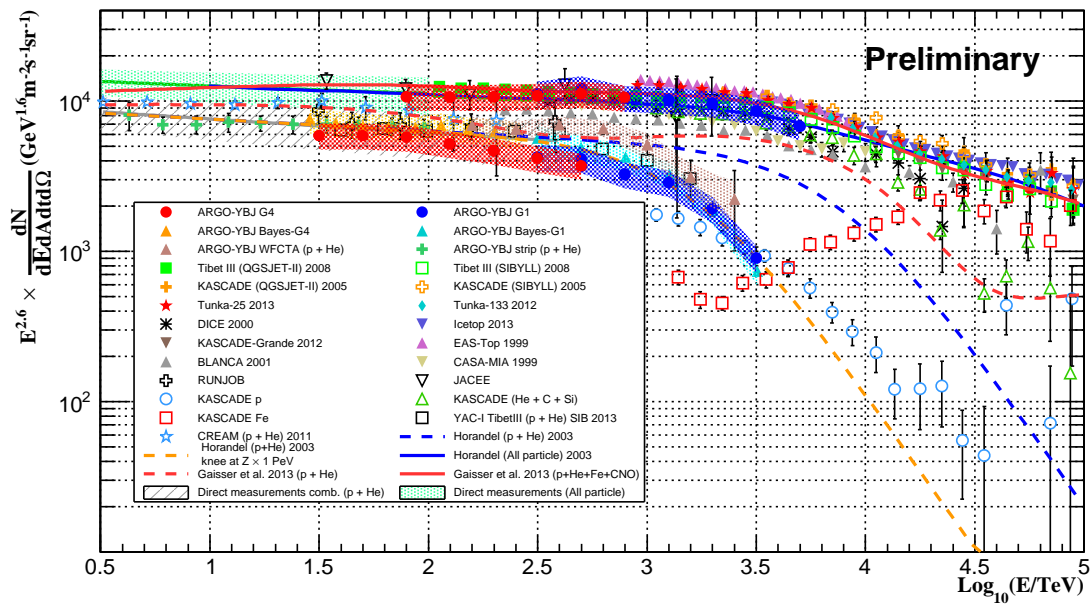} }
\caption{All particle and light (p+He) component energy spectra of primary CR measured by ARGO-YBJ 
and compared to different experimental results. }
\label{fig:all} 
\end{figure*}
%%%%%%%%%%%%%%%%%%%%%%%%%%%%%%%%%%%%%%%%%%%%%%%%%%%%%%%%%%
%

\noindent
All three different analyses are consistent with low energy (direct) measurements and show
a clear evidence for a bending at larger energies but below 1$\,$PeV.
With respect to a single power-law with a spectral index -2.62 the deviation is observed at a level of about 6 s.d.
The results obtained with the two analysis of RPC charge readout data (label 1 and 2 in the previous list) are in fair agreement. 
They also agree with the ARGO-YBJ/WFCTA hybrid measurement within systematic uncertainties and the possible 
difference in the energy scale. 
For comparison, the parametrizations of the light-component provided by \cite{horandel} 
and \cite{gst} are shown by the blue and red dashed lines, respectively. 
A H\"orandel-like spectrum with a modified knee at Z$\times$1 PeV (a factor four lower in energy than in the
original formulation) is also shown for comparison.
Finally ARGO-YBJ results are compared to a compilation of several other measurements in Fig.\ref{fig:all}.

\section{Conclusions}

The CR spectrum has been studied by the ARGO-YBJ experiment in a wide energy range (TeVs$\, \to \,$ PeVs) . 
This study is particularly interesting because not only it allows a better 
understanding of the so called 'knee' of the energy spectrum and of its origin, but also provides 
a powerful cross-check among very different experimental techniques.
The all-particle spectrum (measured in the energy range 100$\,$TeV - 3$\,$PeV) is in good agreement 
with both theoretical parametrizations and previous measurements, making us confident about the 
selection and reconstruction procedures.
The light-component (i.e. p + He) has been measured with high resolution up to about 5 PeV. 
The ARGO-YBJ preliminary result is in agreement with direct measurements and then show a clear indication 
of a bending below 1$\,$PeV. Improvements of event selection with the full statistics and a complete 
analysis of systematic uncertainties is currently under way.

\end{document}